\documentclass[aps,preprint,epsfig,superscriptaddress,rotate]{revtex4}
\usepackage{graphicx}
\usepackage{bm}
\usepackage{setspace}
  \usepackage{setspace}

\def\eg{{e.g.\ }}
\def\ie{i.e.\ }
\def\PD#1#2{\frac{\partial #1}{\partial #2}}    
\newcommand\HAH {\hat{H}}

\def\k #1{|#1\rangle}
\def\b #1{\langle #1|}

\def\mrm #1{\mathrm{#1}}

\def\mrm #1{\mathrm{#1}}
\def\k #1{|#1\rangle}
\def\b #1{\langle #1|}

\input epsf
\begin{document}

\title{Effects of initial coherence on distinguishability of pure/mixed
states and chiral stability in an open chiral system}
\author{Heekyung Han}
\email[E--mail address: ]{hhan0410@gmail.com}
\affiliation{Department of Chemistry, Queen¡¯s University,
Kingston, Ontario K7L , Canada}
\author{David M. Wardlaw}
\email[E--mail address: ]{dwardlaw@uwo.ca}
\affiliation{Department of Chemistry, Queen¡¯s University,
Kingston, Ontario K7L , Canada}
\affiliation{Department of Chemistry, University of Western Ontario,
London, Ontario N6H 5B7, Canada}

\date{\today}

\begin{abstract}
We examine how initial coherences in open chiral systems affect
distinguishability of pure versus mixed states and purity decay.
Interaction between a system and an environment is modeled by a
continuous position measurement and a two-level approximation is
taken for the system. The resultant analytical solution is
explored for various parameters, with emphasis on the interplay of
initial coherences of the system and dephasing rate in determining
the purity decay and differences in the time evolution of pure vs.
mixed initial states. 
process. Implications of the results on several fundamental
problems are noted.
\end{abstract}
\maketitle

\vspace{0.2in} \noindent

\pagebreak

\section{Introduction}\label{intro}
The quantum principle of parity conservation prohibits the presence
of the left- and right-handed states of chiral molecules as true
stationary states, since the Hamiltonian and the parity operator
commute and thus the eigenfunctions of the parity-invariant
Hamiltonian should show parity symmetry in the absence of energy
degeneracy \cite{sakurai}. However, some chiral molecules are
observed to be stable for an enormously long time despite their lack
of parity symmetry.  Theoretical efforts to resolve this conundrum
date back to Hund's proposal in  the early days of quantum-mechanics
\cite{hund}.  He formulated a description of the stability of chiral
molecules in terms of  quantum tunneling in a symmetrical
double-well potential, where  exceedingly long tunneling times
through the inter-well potential barrier were responsible for the
stability. However, one can show that the tunneling time estimates
in observed stable chiral molecules have a wide range, from
sub-picosecond to the age of the universe. On the other hand,
there have been attempts to employ the interaction of molecules with
the environment, which leads to decoherence, and thus to a
suppression of coherent tunneling oscillations, to explain the
stability of chiral molecules. Intermolecular collisions
\cite{collision}, interaction with photons \cite{photon}, and with
phonons \cite{phonon} have been introduced as the physical origin of
the dephasing process.

Another aspect  of stabilization of chiral molecules is the experimental
realization of the superpositions of stable chiral states (so-called
Schr\"{o}dinger's cat state). Several proposals have been suggested for
the preparation and detection of the  superposed chiral states  utilizing
phase-controlled ultrashort laser  pulses \cite{harris_cina, romero}, and
growing attention has been directed towards using lasers to manipulate
the molecular chirality \cite{control_brumer, control_ohtsuki}. Approaches
such as these which rely on the interaction between an external electric
field and a molecule have become more promising with recent advances in
laser technology. Considering that the decoherence effect is crucial to
experimental realizations of coherent superposition of chiral
states, it is thus of great interest to understand the coherent
tunneling dynamics of chiral systems coupled to external
environments more thoroughly.

In this Letter, we study the role of initial coherences, in an open
chiral system, on the distinguishability of initially pure versus
initially mixed states as measured by population difference between
the chiral states, and the vulnerability to decoherence as measured
by the purity decay. The influence of an environment on the system
is modeled by continuous position measurements, an approach often
used in the studies of  the quantum Zeno effect \cite{qzeno} and in
the studies of decoherence effects in chemical reactions
\cite{hhan_jcp}. In particular, Milburn and his coworkers
\cite{deco_model, deco_model2} studied a model of continuous
position measurements on a quantum system in a double-well
potential, and within a two-level approximation their resultant
master equation is identical to ours. However, their goal was the
demonstration of the quantum Zeno effect in certain parameter
regimes and thus an analytical solution was obtained only for an
initially localized state. In contrast, our goal is the
identification of the roles of the initial coherences and the
subsequent decoherence in the distinguishability of pure vs. mixed
chiral states and the stabilization (or loss) of the chirality.
Accordingly we have obtained an analytical solution for an arbitrary
initial condition and explored this solution for a variety of
initial conditions and dephasing rates. Our results provide some
answers to three fundamental questions arising in open chiral
systems: 1) Can we distinguish a pure state from a  mixture with
same populations as those of the pure state? 2) Why are some chiral
systems stable for a very long time? 3) Why are superpositions of
stable chiral states not observed experimentally? Distinguishability
between the initially pure state and the initially mixed state is
measured by a population difference between the two chiral states.
The stability of the system is measured by the purity, which is
equal to 1 only for a pure state and decays as coherences are lost
during an interaction between the system and the environment.

\section{Chiral systems in the presence of dephasing}\label{main}
\subsection{Formalism (a two-level approximation)}\label{formalism}
Consider chiral molecules in a symmetric double-well potential
interacting with a bath and assume that the effects of collisions
between the system and the environment can be modeled by the
continuous measurement of the system coordinate, say $\hat x$
\cite{bath_model,bath_model2}. Then the system density operator
$\hat\rho$, which is obtained by tracing the total density operator
over the environment, obeys the master equation \cite{hhan_jcp},
\begin{eqnarray}
\PD {\hat\rho}{t}=-\frac {i}{\hbar}[\hat
H_0,\hat\rho]-\Gamma[\hat{x},[\hat{x},\hat\rho]].
\label{liouville}\end{eqnarray} Here $\HAH_0$ is the system
Hamiltonian and $\Gamma$ is a coupling strength between the system
and the bath.  The second term in this equation, a multiplication
of $\Gamma$ and the double commutator, represents the
environmental effect on the system, and is expected to destroy
coherences between the eigenstates of the position measurement
operator, $\hat{x}$.  This loss of coherence, \ie, decoherence,
has been argued as the essential ingredient to bridging the gap
between classical and quantum descriptions of a microscopic system
coupled to its environment \cite{decoherence_review2}.

To invoke a two-level approximation, we assume that for an
isolated system the potential barrier is very high  or
equivalently the energy of the system is sufficiently low that the
higher energy levels are not involved in the dynamics. In
addition, since this is an open system, we should point out that
the position measurement of the system is known to cause its
energy and its energy width to increase over time \cite{hhan_jcp,
gallis}. The system energy and energy width will eventually become
greater than the potential barrier, leading to the breakdown in
the validity of the two-level approximation. The stronger $\Gamma$,
the shorter the time during which the two-level approximation will
be valid.

Within the two-level approximation,  we need only the two lowest
energy eigenstates, $\k +$ (even parity; ground state) and $\k -$
(odd parity) with energies $E_+$ and $E_-$. Alternatively,  we can
consider the two localized states $\k L$ and $\k R$, found in the
left- and right-hand wells, respectively. By convention, these
states are given by $\k L =\frac{1}{\sqrt{2}}(\k + +\k -) \,\,\,$
and $\k R =\frac{1}{\sqrt{2}}(\k + -\k -)$.  The free Hamiltonian
readily takes the form
$\hat H_0=E_+\k + \b +\,\, + \,\,E_-\k - \b-
=E_m(\k R \b R + \k L \b L)-\delta(\k L \b R + \k R \b
L)$ 
where $E_m=\frac{(E_+ +E_ -)}{2}$ and $\delta =\frac{(E_-
-E_+)}{2}\equiv \hbar \omega$ with a tunneling frequency $\omega$.
For simplicity, choosing the energy origin to correspond to $E_m=0$,
\ie, $E_+=-E_-$ and using the usual Pauli matrices $\hat{\sigma_x}$,
$\hat{\sigma_y}$, $\hat{\sigma_z}$ in the $\k L$ $\k R$ basis, $H_0$
becomes $H_0=-\delta\hat{\sigma_x}$.
Using the two-level approximation and the evenness and oddness of
$\k +$ and $\k -$, respectively, we derive the position
measurement operator as
$\hat{x}= \k +\b + \hat{x} \k - \b - +{\mrm{H.c}}
=x_{\mrm{avg}}\hat{ \sigma_{z}}
\approx x_{\mrm{min}} \hat{ \sigma_{z}}$, 
where  ${\mrm{H.c}}$ denotes the Hermitian conjugate and $\pm
x_{\mrm{min}}$ are the positions of the double-well minima, which for a
deep well are very close to the average positions of the right and left
wave functions (denoted by $\pm x_{\mrm{avg}}$). Finally the master
equation for the two-level system becomes \cite{deco_model}
  \begin{eqnarray}
\PD {\hat\rho}{t}=\frac {i}{\hbar}[\delta
\hat{\sigma_x},\hat\rho]-\gamma[\hat\sigma_z,[\hat\sigma_z,\hat\rho]],
\label{liouville12l}\end{eqnarray} with $\gamma\equiv\Gamma
x_{\mrm{min}}^2$.   This equation gives the evolution of the
tunneling two-level system subjected to a continual measurement of
the system coordinate $\hat{x}$ as represented by the Pauli matrix
$\hat\sigma_z$. The second term in this equation represents a
measurement of  a population difference between two states $\k L $
and $\k R$, and leads to decay of the coherences between the $\k L
$ and $\k R$ states, \ie, of the off-diagonal matrix elements,
$\rho_{LR}$.   Note that here $\gamma$ is given by ``a
multiplication of $\Gamma$ by $x_{\mrm{min}}^2$'', which means that
the magnitude of $\gamma$ is determined by the interaction
strength (between the system and the bath) and the system
potential as well. Thus within the two-level approximation,
$\gamma$ can be increased by increasing the interaction strength
or by increasing the distance between two wells of the potential.
It is intuitive that for a system composed of two wavepackets that
are initially spatially separated and subjected to an environment
destroying coherences between these two wavepackets, the larger
the distance between two wavepackets and the larger the
interaction strength between the system and the environment, the
faster coherences between two wavepackets are lost \cite{qzeno}.

It is  also noteworthy that this master equation is identical to
that resulting from a two-level atom driven by a field undergoing a
pure dephasing \cite{qoptics} in quantum optics, as previously
mentioned in Ref. \cite{deco_model,deco_model2}.  This equivalency
emerges where the following replacements are made: $\k
{L,R}\rightarrow$ the two energy eigenstates, $\omega \rightarrow$
Rabi-frequency, and $\gamma\rightarrow$ pure dephasing rate.

The density-matrix elements of the states $\k L$
and $\k R$, $\rho_{LL}$, $\rho_{RR}$, and  $\rho_{LR}$ obey the
following set of equations:
\begin{eqnarray}
  \PD{\rho_{LL}}{t}&=&-\mrm{Im}(2\omega\rho_{RL}),\label{rhol}\\
  \PD{\rho_{RR}}{t}&=&\mrm{Im}(2\omega\rho_{RL}),\label{rhor}\\
  \PD{\rho_{RL}}{t}&=&-4\gamma\rho_{RL}+i\omega(\rho_{LL}-\rho_{RR})
  \label{rhorl}
\end{eqnarray}
It is convenient to switch to the  Bloch-vector representation by
defining  the real quantities, $X$, $Y$, and  $Z$ as
$X=2{\mrm{Re}}\,(\rho_{LR})$, $ Y=2{\mrm{Im}}\,(\rho_{LR})$, and
$Z=\rho_{RR}-\rho_{LL}$ \cite{two_level_atoms}.  (In regard to
relationship of $X$, $Y$, and  $Z$ to  coherence and localization of
the system, it is noteworthy that for a pure state, since
$X^2+Y^2+Z^2=1$,  the localized state, say $\k R$, corresponds to
$Z^2= 1$, \ie, $X^2+Y^2= 0$, and the  strongly delocalized state
corresponds to $Z^2\approx 0$, \ie, $X^2+Y^2\approx 1$.)  Rewriting
Eqs. (\ref{rhol}) to (\ref {rhorl}) in terms of $X$, $Y$, and $Z$:
\begin{eqnarray}
\frac{dX}{dt}&=&-4\gamma X,\label{X}\\
\frac{dY}{dt}&=&-4\gamma Y + 2 \omega Z,\label{Y}\\
\frac{dZ}{dt}&=&-2\omega Y.\label{Z}
\end{eqnarray}
The resultant equations are of standard form \cite{torrey}. Thus the study
of chiral stability is reduced to the solution of the $3 \times 3$
dynamical system which depend upon two parameters $\omega$ and $\gamma$.
Note that the longtime steady-state solution of Eqs.(\ref{X})-(\ref{Z}) is
$X = Y = Z = 0$ as $t\rightarrow \infty$. It can be obtained by setting
$dX/dt = dY/dt = dZ/dt = 0$. This stationary state, which is reached
regardless of initial conditions, system parameters, or dephasing rate,
represents a fully mixed state, with equal populations in the right and
left wells and with no remaining coherence between them.

With an arbitrary initial state [$X = X_0, Y = Y_0, Z = Z_0$], the general
solution is, for $\omega>\gamma$ (tunneling dominant region),
\begin{eqnarray}
X&=&X_0e^{-4\gamma t} \label{solu},\\
Y&=&-\frac{e^{-2\gamma t}}{2\omega}[\omega
Y_0\{-2\cos(st)+\frac{4\gamma}{s}\sin(st)\}-Z_0(
s+\frac{4\gamma^2}{s})\sin(st)]\label{solv},\\
Z&=&e^{-2\gamma t}[Z_0 \{\cos
(st)+\frac{2\gamma}{s}\sin(st)\}-\frac{2\omega Y_0}{s} \sin
(st)]\label{solw},
\end{eqnarray}
with $s=2\sqrt{\omega^2-\gamma^2}$. The population of the system in either
well ($\rho_{RR}=\frac{1+Z}{2}, \rho_{LL}=\frac{1-Z}{2}$) oscillates at the
modified frequency $s$, that is determined by relative size of the
dephasing rate $\gamma$ and the so-called tunneling frequency $\omega$,
under an exponentially decaying envelope. The introduction of decoherence
increases the period of the oscillation (since the frequency $s$
decreases as $\gamma$ increases from zero towards $\omega$) while
simultaneously suppressing the amplitude via the decay factor $e^{-\gamma
t/2}$ in the tunneling dominant region.

On the other hand, for $\gamma>\omega$ (dephasing dominant region), the
general solution is obtained by replacing $\sin(st)$ by $\sinh(\tilde{s}t)$
and $\cos(st)$ by $\cosh(\tilde{s}t) $ with $\tilde{s} = 2
\sqrt{\gamma^2-\omega^2}$ in Eqs. (\ref{solu})-(\ref{solw}):
\begin{eqnarray}
X&=&X_0e^{-4\gamma t} \label{solu2},\\
Y&=&-\frac{e^{-2\gamma t}}{2\omega}[\omega
Y_0\{-2\cosh(\tilde{s}t)+\frac{4\gamma}{\tilde{s}}\sinh(\tilde{s}t)\}-Z_0(
\tilde{s}+\frac{4\gamma^2}{\tilde{s}})\sinh(\tilde{s}t)]\label{solv2},\\
Z&=&e^{-2\gamma t}[Z_0 \{\cosh
(\tilde{s}t)+\frac{2\gamma}{\tilde{s}}\sinh(\tilde{s}t)\}-\frac{2\omega
Y_0}{\tilde{s}} \sinh (\tilde{s}t)]\label{solw2}.
\end{eqnarray}
In this regime for an initially localized state, say $\k R$, the
system approaches the longtime steady-state monotonically without
any oscillations present. Ref.\cite{deco_model} demonstrated the
quantum Zeno effect in this regime: when the measurement strength
exceeds some threshold, the system initially prepared in one well
remains stuck on a time scale  of the measurement strength. If we
attempt to apply this observation to our open chiral system, then we
may say that for an initially chiral system the time over which the
chirality is preserved will increase proportionally when $\gamma$
increases beyond a certain threshold (here $\omega$). Indeed, this
is what we observe in the case with an initially localized state for
a dephasing dominant region in Sec.\ref{purity decay}. Further,
considering that previous studies focused  on manifestation of the
quantum Zeno effect for initially localized states, it would be
interesting to explore whether and how this quantum Zeno effect,
namely, the suppression of a transition (tunneling), exists in the
case of an initially delocalized state; this will be reported in
Sec.\ref{purity decay}.

\subsection{Distinguishability}\label{distinguishability}
In this subsection, we examine how the initial coherences affect, over
time and under the influence of decoherence, the distinguishability
between the initially pure state and the mixed state. These comparisons
require selection of initial conditions for the pure state
$(X_0^{\mrm{P}},Y_0^{\mrm{P}},Z_0^{\mrm{P}})$ and the mixed state
$(X_0^{\mrm{M}},Y_0^{\mrm{M}},Z_0^{\mrm{M}})$. The most meaningful
comparison arises from assigning the same initial population of $\k L$
and $\k R$ in the initial ($t=0$) pure and mixed states, \ie
$Z_0^{\mrm{P}}=Z_0^{\mrm{M}}$, but choosing the degree of initial
coherence to be zero for the mixed state (maximum randomness
scenario), \ie $X_0^{\mrm{M}}=Y_0^{\mrm{M}}=0$, while varying
$Y_0^{\mrm{P}}$ for the pure state. Note that $X_0^{\mrm{P}}$ is
determined by $\{X_0^{\mrm{P}}\}^2+\{Y_0^{\mrm{P}}\}^2
+\{Z_0^{\mrm{P}}\}^2=1$, and $X_0^{\mrm{P}}$ will not affect the
distinguishability as measured by $Z$ since $X$ and $Z$ are decoupled
(see Eqs. (\ref{X})-(\ref{Z})). As easily can be seen from
Eqs.(\ref{solw}) and (\ref{solw2}), the difference between the
time-evolution of $Z$ of the initially pure state and that of the
corresponding mixed state, $\Delta Z (\equiv Z^{\mrm{P}} -
Z^{\mrm{M}})$ is given by:
\begin{eqnarray}
\Delta Z(t)= -\frac{2\omega Y_0^{\mrm{P}}
  }{s} e^{-2\gamma t}\sin (st)\,\, &{\mrm{for}}&
\omega>\gamma,\label{z_diff1}\\
\Delta Z(t) = -\frac{\omega Y_0^{\mrm{P}} }{\tilde{s}}
[e^{(\tilde{s}-2\gamma )t}-e^{-(\tilde{s}+2\gamma )t}]\,\,
&{\mrm{for}}& \omega<\gamma.\label{z_diff2}
\end{eqnarray}
From the above results one can discuss  the distinguishability
between the initially pure state and the corresponding initially
random mixture by measuring the population difference $Z$ in the
presence/absence of the bath effect destroying the right-left
coherences. Note that the distinguishability is not affected by
$X_0^{\mrm{P}}$ as expected and that it depends strongly on the
magnitude of $Y_0^{\mrm{P}}$. Clearly, if $Y_0^{\mrm{P}}=0$, $\Delta
Z(t)=0$ for all time and for all $\gamma$. That is, if
$Y_0^{\mrm{P}}=0$, one can not distinguish the initially pure state
from the corresponding random-mixed state by measuring $Z$,
regardless of the decoherence effect. However, if
$Y_0^{\mrm{P}}\neq 0$ the distinguishability does exist, although
the decoherence effect tends to extinguish it over time because the
decoherence forces both the initially pure state and the
corresponding mixture to the maximally randomized stationary state
(equal $\k L$ and $\k R$ populations without any remaining
coherences between the two states at long times). It is noteworthy
that our result reproduces, in the limit of no bath and no
time-evolution of the system, the work by Harris and his coworkers
\cite{harris_distinguish}. They assumed essentially instantaneous
measurements, occurring on a time scale short compared to that for
tunneling and dephasing, and, in the framework of the wavefunction,
showed that no parity sensitive experiment can measure the
difference between them if the states $\k L $ and $\k R$ are taken
to be real for the pure state, which corresponds to the case of
$Y_0^{\mrm{P}}=0$.

The behavior of the distinguishability also depends on the relative
size of the dephasing rate and the tunneling frequency. For the
tunneling dominant region ($\omega>\gamma$), $\Delta Z(t)$ shows
oscillations that are exponentially damped with time, and whose
amplitude at a given time tends to increase with increasing $\omega$
and $Y_0^{\mrm{P}}$.  On the other hand, for the dephasing dominant
region ($\omega<\gamma$), $|\Delta Z(t)|$ shows a simpler behavior:
it increases  up to maximum at $t=t_{\mrm{max}}$ from zero at $t=0$,
and then decreases to zero as time increases. Here $t = t_{\mrm{max}} =
\frac{1}{2\tilde{s}}\ln[\frac{\tilde{s}+2\gamma}{-\tilde{s}+2\gamma}]$
is obtained such that $\frac{d (\Delta Z(t))}{dt}|_{t=t_{\mrm{max}}} = 0$.

\begin{figure}[htbp]
\centerline{\hbox{\epsfxsize=6.0in \epsfbox{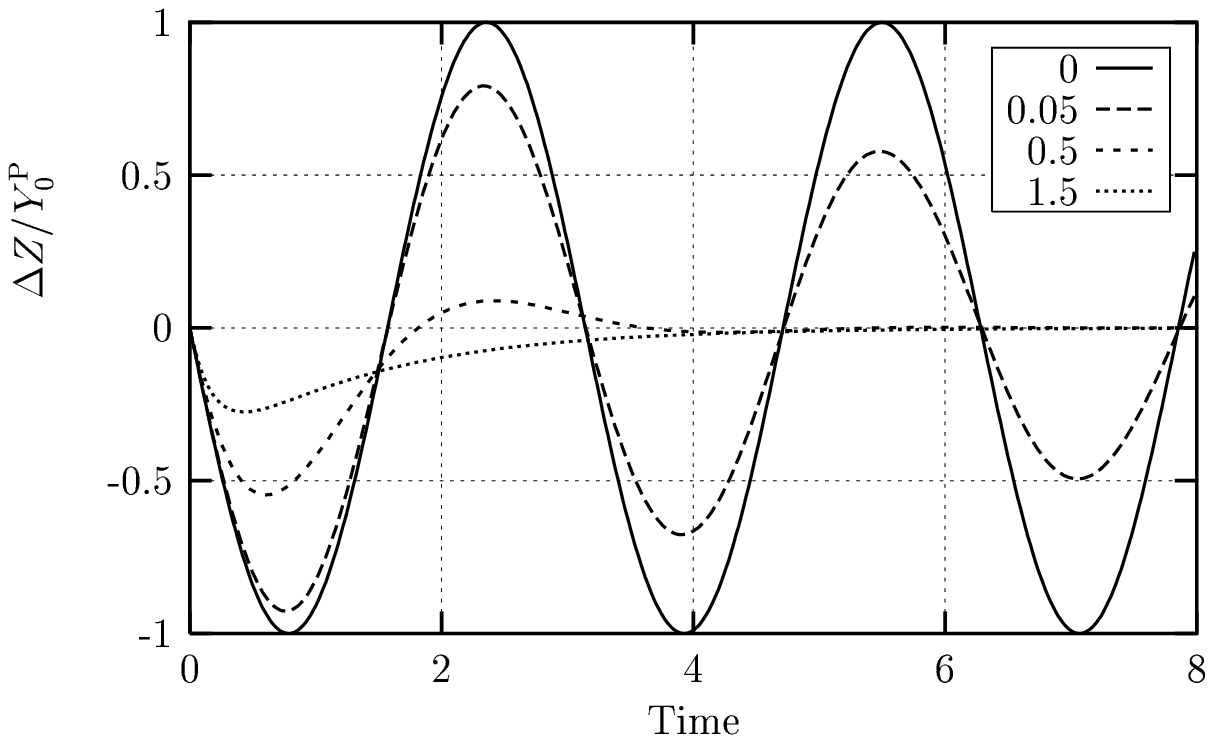}}}
\centerline{\hbox{\epsfxsize=6.0in \epsfbox{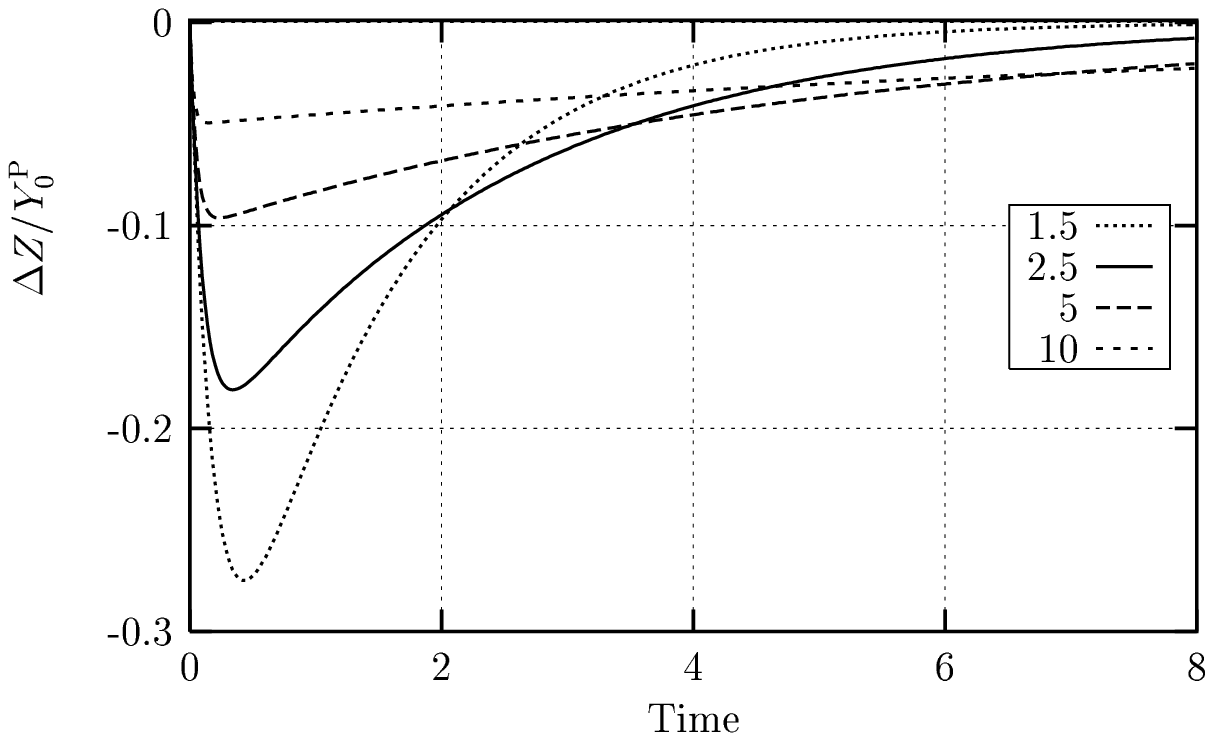}}}
\caption{$\Delta Z (t)
/ Y_0^{\mrm{P}}$ vs. time for various  $\gamma$ shown inside the
box. 
Upper panel: $\gamma=0$ (dephasing-free case), $\gamma=0.05, \,0.5$
(tunneling dominant region), and $\gamma=1.5$ (dephasing dominant
region). Lower panel: all are in dephasing dominant region. All the
variables are in dimensionless units. }
  \label{deltaz}
  \end{figure}

Typical behavior of $\Delta Z(t)$ for several values of $\gamma$ for
a given value of $\omega$ (here chosen as 1) is shown in Fig.
\ref{deltaz}.  For $\omega > \gamma$ (tunneling dominant region,
here $\gamma < $ 1), as  $\gamma$ increases, the period of the
oscillation increases and  the amplitude  decays faster, leading
$\Delta Z$ to a faster approach toward zero ($\Delta Z=0$  means no
distinguishability).  For example, for $\gamma=0.5$ [upper panel],
upon the measurement of population  after $t=4$, one can not tell
whether the system was initially in the pure state or in the
corresponding mixture. On the other hand, for $\omega < \gamma$
(dephasing dominant region, here $\gamma > $ 1),  after hitting the
minimum point,  $\Delta Z /Y_0^{\mrm P}$ just increases
monotonically towards 0. It is noteworthy that, as $\gamma$
increases, the distinguishability disappears more slowly, while the
maximum of $|\Delta Z /Y_0^{\mrm P}|$ gets smaller and  is obtained
at earlier time, \ie,  $t_{\mrm{max}}$ gets smaller.  For instance,
compare the case $\gamma=1.5$ and $\gamma=10$ in the lower panel.
For the case $\gamma=1.5$, $|\Delta Z /Y_0^{\mrm P}|=0.275$ at
$t=t_{\mrm{max}}=0.42$, and almost zero at $t=8$, indicating no
distinguishability for $t>8$, while, for the case $\gamma=10$,
$|\Delta Z /Y_0^{\mrm P}|=0.0494$ at $t=t_{\mrm{max}}=0.14$, and
still nonzero as 0.0226  (roughly half of maximum  $|\Delta Z
/Y_0^{\mrm P}|$)  at $t=8$, indicating the long-time surviving
distinguishability.  This observation suggests that  considerably
strong dephasing may  unexpectedly afford means to observe the
long-time surviving distinguishability.

\subsection{Decoherence (purity decay)}\label{purity decay}
In this subsection we examine how initial coherences affect the
decoherence process. In order to measure coherence decay, \ie,
decoherence, we use the purity, $\varsigma$ defined as
\cite{purity}
$\varsigma\equiv \mathrm{Tr}(\hat{{\rho}}^{2})$,
where $\mathrm{Tr}$ denotes a trace over the system of interest.
Note that the decoherence process does not preserve the purity of
the state, that is, the purity of the resultant mixed state
becomes less than 1, while that of the pure state is 1.   We first
sketch some features of the effects of initial coherences and
dephasing rate on the purity decay (decoherence process), and then
present the results of numerical calculations.

Noting that for a two-level system purity is given by
$\varsigma=\frac{1}{2}(X^2+Y^2+Z^2+1)$ in the Bloch-vector representation
and  using Eqs. (\ref{X})-(\ref{Z}), one can obtain
\begin{eqnarray}
\frac{d\varsigma}{dt}=-4\gamma(X^2+Y^2) \label{puritydt}
\end{eqnarray}
From Eq.(\ref{puritydt}) one can expect that at fixed $\gamma$,
for short times, purity decay gets faster with increasing $X_0$
and/or $ Y_0$ under the restriction $X_0^2 + Y_0^2 + Z_0^2\le 1$ (the
equality holds only for the pure state). However, for the long
time region the effects of $X_0$ and $Y_0$ on the purity decay
are different due to  different  time-evolutions of $X$ and $Y$
(see Eqs.(\ref{solu})-(\ref{solw}) and Eqs.(\ref{solu2})-(\ref{solw2})).
While $X$ exponentially decays to zero (see Eq. (\ref{X})), $Y$'s
behavior over time is more complicated due to its coupling to $Z$ as
seen in Eqs.(\ref{Y}) and (\ref{Z}). It is notable that purity decays
invariantly with regard to $X_0\rightarrow - X_0$ regardless of time
and dephasing rates, but not with regard to $Y_0\rightarrow - Y_0$.
On the other hand, from Eq.(\ref{puritydt}) at a given initial
condition purity is expected to decay faster with increasing dephasing
rates $\gamma$ for short times such that $X\approx X_0$ and $Y \approx
Y_0$. However, this tendency may change in the long time region,
depending on conditions. Details of the effects of initial coherences,
$(X_0,Y_0)$ and dephasing on the purity decay are explored below.

All numerical results are for initially pure states. One could also
examine purity decay in initially mixed states but the initially
pure state maximizes the influence of initial coherences and is
inherently more interesting in terms of applications. We consider
sets of initially pure states differing in the degree of R and L
admixture. That is, $\varsigma(t=0) = 1$ and thus $X_0^2 + Y_0^2 +
Z_0^2=1$. Here $Z_0$ is chosen as $1-\sqrt{X_0^2 + Y_0^2}$. Also
for all the cases examined $\omega$ is chosen as 1, and all the
variables are in dimensionless units.

\begin{figure}
\begin{center}
\begin{tabular}{cc}
{\hbox{\epsfxsize=3.2in \epsfbox{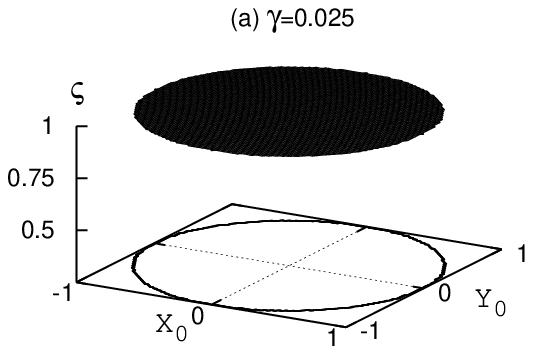}}}&
{\hbox{\epsfxsize=3.2in \epsfbox{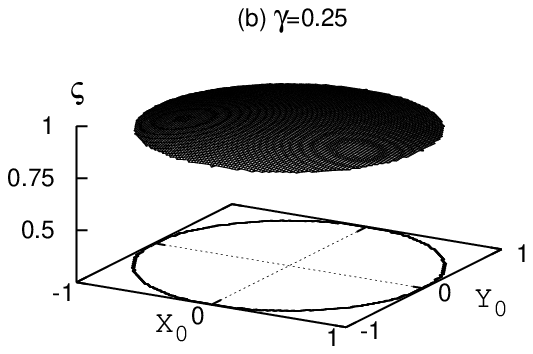}}}\\
{\hbox{\epsfxsize=3.2in \epsfbox{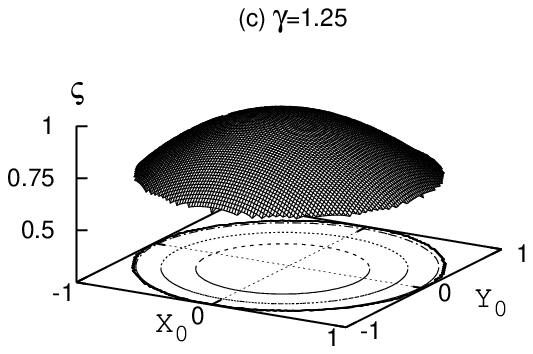}}}&
{\hbox{\epsfxsize=3.2in \epsfbox{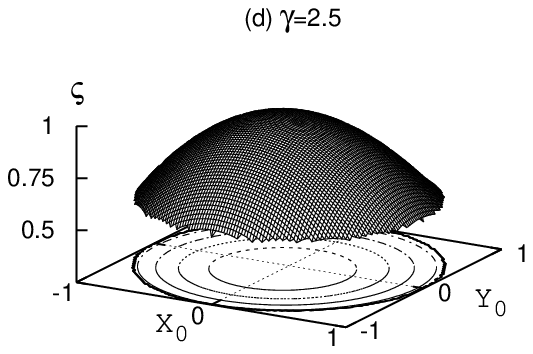}}}\\
{\hbox{\epsfxsize=3.2in \epsfbox{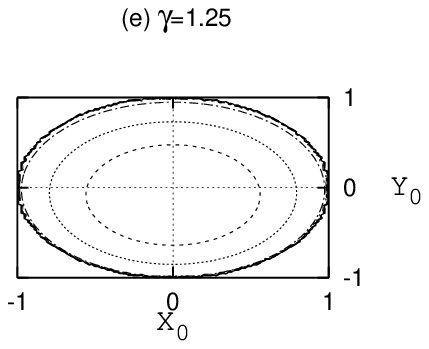}}}&
{\hbox{\epsfxsize=3.2in \epsfbox{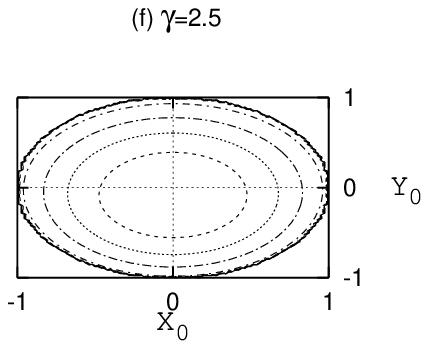}}}\\
\end{tabular}
  \caption{Purity  $\varsigma$
vs. initial coherences $(X_0,Y_0)$ at time $t=0.1$ for various
dephasing rates $\gamma$: (a) $\gamma=0.025$, (b) $ \gamma=0.25$,
(c) $\gamma=1.25$, and (d)  $\gamma=2.5$ ((e) and (f) are
2-dimensional contour plots for (c) and (d), respectively).  All the
variables are in dimensionless units.
  }
\label{time_01}\end{center}\end{figure}

\begin{figure}
\begin{center}
\begin{tabular}{cc}
{\hbox{\epsfxsize=3.2in \epsfbox{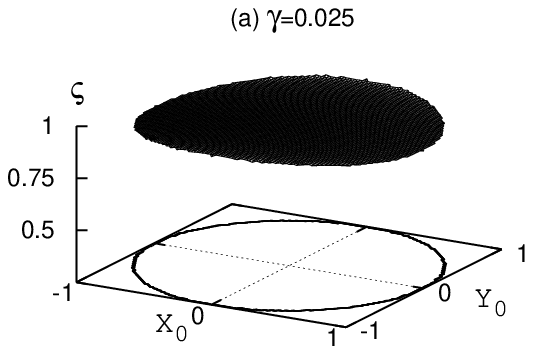}}}&
{\hbox{\epsfxsize=3.2in \epsfbox{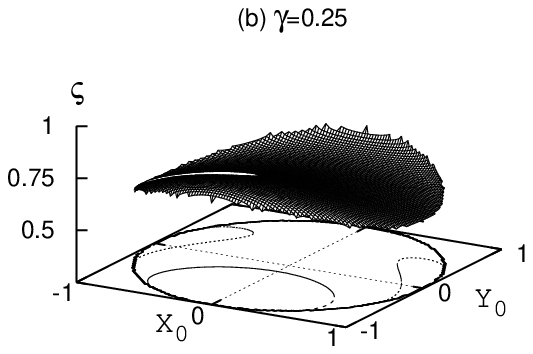}}}\\
{\hbox{\epsfxsize=3.2in \epsfbox{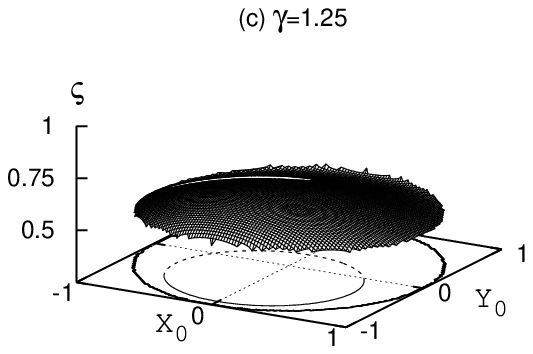}}}&
{\hbox{\epsfxsize=3.2in \epsfbox{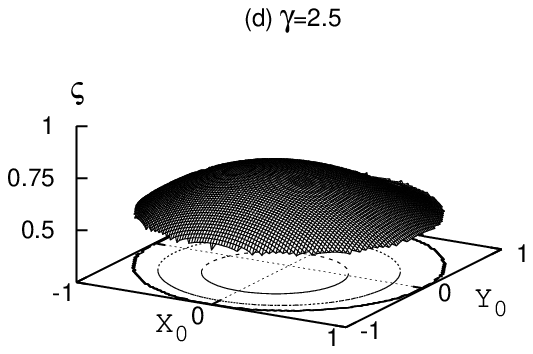}}}\\
\end{tabular}
\caption{ As in Fig. \ref{time_01} but  at time $t=1$.
   }
\label{time_1}\end{center}\end{figure}

Figures \ref{time_01} and \ref{time_1} show purity as a function of
initial coherences $(X_0,Y_0)$ with various dephasing rates spanning
the tunneling dominant region ($\gamma<1$, (a) and (b)) to the
dephasing dominant region ($\gamma>1$, (c) and (d)) at times $t=0.1$,
and 1, respectively. Several interesting observations are in order.
First, one can easily see that for a given time and dephasing rates,
purity is symmetric about $X_0=0$ at a fixed $Y_0$, but not symmetric
about $Y_0=0$ at a fixed $X_0$. This implies that as $|X_0|$ increases
at a fixed $Y_0=0$, purity decays faster for a given time and dephasing
rate, but, however, as $|Y_0|$ increases at a fixed $X_0=0$, decay
of purity may be suppressed. Figs. \ref{time_1} (b) and (c) capture this
feature very well. For instance, in Fig.\ref{time_1} (c) one can easily
see that, as $Y_0$ increases in a negative direction from $Y_0=0$ with a
fixed $X_0=0$, purity increases up to some point (\ie, purity increases
with increasing $|Y_0|$), and then decreases. Second, for very short
times such as $t=0.1$ [Fig. \ref{time_01}], purity tends to decay
faster, as $\gamma$ increases and as $X_0^2+Y_0^2$ increases. This
agrees well with the prediction of Eq. (\ref{puritydt}), which is
$\varsigma(t)\approx 1-4\gamma(X_0^2 + Y_0^2)t$ for very short times
such that $X\approx X_0$ and $Y\approx Y_0$.  Also note that this
proportionality to $X_0^2+Y_0^2$ gets distorted for the larger
$\gamma$; the circle in the contour of the purity gets distorted in
Figs. \ref{time_01} (c), (e) and (d), (f). This is because
increasing $\gamma$ accelerates changes of $X, Y$ from  $X_0, Y_0$.
Third, on the other hand, for later times (see Fig. \ref{time_1}),
and in the region of larger initial coherences ($X_0^2+Y_0^2\approx
1$), purity still tends to decay faster with increasing $\gamma$,
but for the smaller initial coherences ($X_0^2+Y_0^2\approx 0$),
purity shows the opposite behavior, with increasing $\gamma$ beyond
$\omega$, \ie, the purity decay starts to be suppressed. That is, if
initially the system is strongly localized in one of the two wells
(smaller initial coherences), then introducing sufficiently large
dephasing (considerably larger than tunneling frequency) will suppress
the decoherence process (racemization), but, however, if initially
the system is strongly delocalized in two wells (larger initial
coherences), then increasing the dephasing rate could accelerate
the decoherence process and the purity decay at a rate faster than
for the initially well-localized system for a given condition.
This different behavior of the system in respect to increasing
$\gamma$, depending on its initial condition, can be understood by
recognizing the role of the environment as destroying the
left-right coherences, and thus inhibiting tunneling of the
system. The tunneling (transition) suppression leads to the
stabilization of the chirality for the initially localized state
(or so-called quantum Zeno effect) and the faster racemization
for the initially delocalized state. Implications of this
interesting result contribute to answers to some of long-standing
fundamental problems in chiral systems: 1) why are the chiral
states of some molecules, once produced, stable for a very long
time? 2) why are superpositions of stable chiral states not
observed? Our result suggests that the system, if it is
initially a well-localized state, will tend to preserve the
chirality in the external environment that destroys the left-right
coherences, provided that the interaction with the environment
dominates the tunneling interaction, while, on the other hand, the
system, if initially a superposition of two chiral states, would
become very quickly a racemic mixture in the presence of the very
same external environment.

\section{summary}\label{summary}
We have obtained an analytical solution for an open chiral system.
The interaction of the system with an environment was modeled as a
continuous position measurement of the system and a two-level
approximation was made for the system. The interplay of the initial
coherences and the decoherence on the distinguishability between
the pure/mixed states, and the racemization/stabilization of the
chirality was clarified. If the two chiral states are initially
taken to be real for the pure state (\ie, $Y_0^{\mrm{P}}=0$), no
distinguishability was shown to exist for all time and the dephasing
rate, as measured by a population difference between the two chiral
states, was essentially zero. Increasing the dephasing rate was
shown to tend to accelerate the distinguishability decay for
$Y_0^{\mrm{P}}\neq 0$,
and the purity decay over time. However, considerably strong
dephasing (beyond a certain threshold) was observed to enable the
long-time survival of distinguishability, and the chirality
stabilization for the initially localized state. The results should
serve as a prototype for understanding the results of a chiral
system under more complicated and realistic conditions, \eg, more
energy levels, more spatial degrees of freedom, and other types of
external environment which give rise to chiral or achiral
interactions.

This work may be extended to a chiral system interacting with lasers.
Recently there has been considerable interest on the preparation/control
of molecular chirality
\cite{harris_cina,romero,control_brumer,control_ohtsuki}. However, the
decoherence effect on these preparation/control scenarios is far from
well understood, especially when the time scales of system dynamics,
and/or environmental effects (energy loss, dephasing) are comparable to
the interaction time scale between the system and the laser field.

\begin{center}
     {\bf Acknowledgements}
\end{center}

It is a pleasure to acknowledge the Natural Sciences and Engineering
Research Council of Canada for financial support in the form of a Discovery
Grant. Authors thank Alexei M. Frolov (UWO) for consultations on aspects of
this manuscript.

\newpage

\end{document}